\documentclass{aa}
\usepackage[utf8]{inputenc}
\usepackage{natbib}
\usepackage{graphicx}
\usepackage{amsmath}
\usepackage{gensymb}
\usepackage{subcaption}
\usepackage{dblfloatfix}
\usepackage{makecell}
\usepackage{ragged2e}
\usepackage[colorlinks=true,allcolors=blue]{hyperref}
\usepackage[scientific-notation=true]{siunitx}
\usepackage{xcolor}
\usepackage{mhchem}

\title{A complete Herbig disk mass survey in Orion}

\author{L. M. Stapper\inst{\ref{inst1}, \ref{inst1.2}} \and M. R. Hogerheijde\inst{\ref{inst1}, \ref{inst2}} \and E. F. van Dishoeck\inst{\ref{inst1},\ref{inst3}} \and A. S. Booth\inst{\ref{inst4}} \and S. L. Grant\inst{\ref{inst3}} \and S. E. van Terwisga\inst{\ref{inst5}}}

\institute{Leiden Observatory, Leiden University, PO Box 9513, 2300 RA Leiden, The Netherlands \label{inst1} \and Max-Planck-Institut für Astronomie, Königstuhl 17, 69117 Heidelberg, Germany \\e-mail: \texttt{lustapper@mpia.de} \label{inst1.2} \and Anton Pannekoek Institute for Astronomy, University of Amsterdam, PO Box 94249, 1090 GE, Amsterdam, The Netherlands \label{inst2} \and Max-Planck-Institut für Extraterrestrische Physik, Giessenbachstrasse 1, 85748 Garching, Germany \label{inst3} \and Center for Astrophysics | Harvard \& Smithsonian, 60 Garden Street, Cambridge, MA 02138, USA \label{inst4} \and Space Research Institute, Austrian Academy of Sciences, Schmiedlstr. 6, 8042, Graz, Austria \label{inst5}}

\date{\today}

\abstract
{ %Context
Disks around intermediate mass stars called Herbig disks are the formation sites of giant exoplanets. Obtaining a complete inventory of these disks will therefore give insights into giant planet formation. However, until now no complete disk survey has been done on Herbig disks in a single star-forming region.
}
{ %Aims
This work aims to obtain the first complete survey of Herbig disks. Orion is the only nearby region with a significant number of Herbig disks (N=35) to carry out such a survey. The resulting dust mass distribution is compared to other dust mass distributions of disks around proto- and pre-main sequence stars in Orion. In addition we ascertain if previous ALMA observations have been biased towards the most massive and brightest Herbig disks.
}
{ %Methods
Using new NOEMA observations of 25 Herbig disks, in combination with ALMA archival data of 10 Herbig disks, results in a complete sample of all know Herbig disks in Orion. Using $uv$-plane analysis for the NOEMA observed disks, and literature values of the ALMA observed disks, we obtain the dust masses of all Herbig disks and obtain a cumulative dust mass distribution. Additionally, six disks with new CO isotopologues detections are presented, one of which is detected in \ce{C^17O}. We calculate the external ultraviolet (UV) irradiance on each disk and compare the dust mass to it.
}
{ %Results
We find a median disk dust mass of 11.7~$M_\oplus$ for the Herbig disks. Comparing the Herbig disks in Orion to previous surveys for mainly T~Tauri disks in Orion, we find that while $\sim50$\% of the Herbig disks have a mass higher than 10~M$_\oplus$, this is at most 25\% for the T~Tauri disks. This difference is especially striking when considering that the Herbig disks are around a factor of two older than the T~Tauri disks. Comparing to the Herbig disks observed with ALMA from a previous study, no significant difference is found between the distributions. We find a steeper (slope of $-7.6$) relationship between the dust mass and external UV irradation compared to that of the T~Tauri disks (slope of -1.3). Comparing our results to a recent SPHERE survey of disks in Orion, we see that the Herbig disks present the largest and brightest disks and have structures indicative of gas-giant formation.
}
{ %Conclusions
Herbig disks are on average more massive compared to T~Tauri disks. This work shows the importance of complete samples, giving rise to the need of a complete survey of the Herbig disk population.
}


\begin{document}

\maketitle

\section{Introduction}
\label{sec:introduction}
Herbig disks are disks around pre-main sequence stars with spectral types of B, A, and F, and stellar masses of 1.5-10~M$_\odot$ with H$\alpha$ indicating ongoing accretion \citep{Herbig1960, Brittain2023}. These disks are the prime formation site of giant exoplanets: directly imaged exoplanets are often found around early spectral type stars \citep{Marois2008, Marois2010, Lagrange2010}, and exoplanet population studies show that the occurrence rate of massive exoplanets is highest around intermediate mass stars \citep[e.g.,][]{Johnson2010, Nielsen2019}. Recent work by \citet{Stapper2022} has shown that the mean mass of Herbig disks is higher compared to disks around lower mass T~Tauri stars. The interpretation of these higher disk fluxes are still being discussed: either as a consequence of massive exoplanets forming in these disks stopping radial drift and keeping the emitting area of the dust large \citep{Stapper2022}, and/or high dust masses causing massive exoplanets to form in these disks \citep[e.g.,][]{GuzmanDiaz2023}. Regardless of interpretation, massive exoplanets are likely forming in these disks.

Some of the most well-known protoplanetary disks are Herbig disk. These millimeter-bright disks display many different types of structures, and are therefore especially favored for in-depth morphological \citep[e.g.,][]{Andrews2018b}, kinematical \citep{Pinte2018, Pinte2019, Izquierdo2022}, and chemical studies \citep[e.g.,][]{Oberg2021, Booth2024}. \citet{Stapper2022} compiled all ALMA data available of Herbig disks within 450~pc, including these well-studied disks. They found a clear increase in the mean dust mass compared to disks in the Lupus and Upper~Sco star-forming regions, which primarily consist of T~Tauri stars, showing that the stellar mass - disk mass relationship extends to the intermediate mass regime. Still, both Herbig and T Tauri disks span the same range of masses, although their distributions are skewed to different averages. Since the T Tauri samples are complete and the Herbig sample of \citet{Stapper2022} is constructed from Herbig disks studied in a variety of ALMA projects (still 64\% complete including all nearby star-forming regions), the question remains if the ALMA coverage of Herbig disks is biased toward well-known and `interesting' objects with higher disk masses.

\begin{figure}[t]
    \centering
    \includegraphics[width=0.45\textwidth]{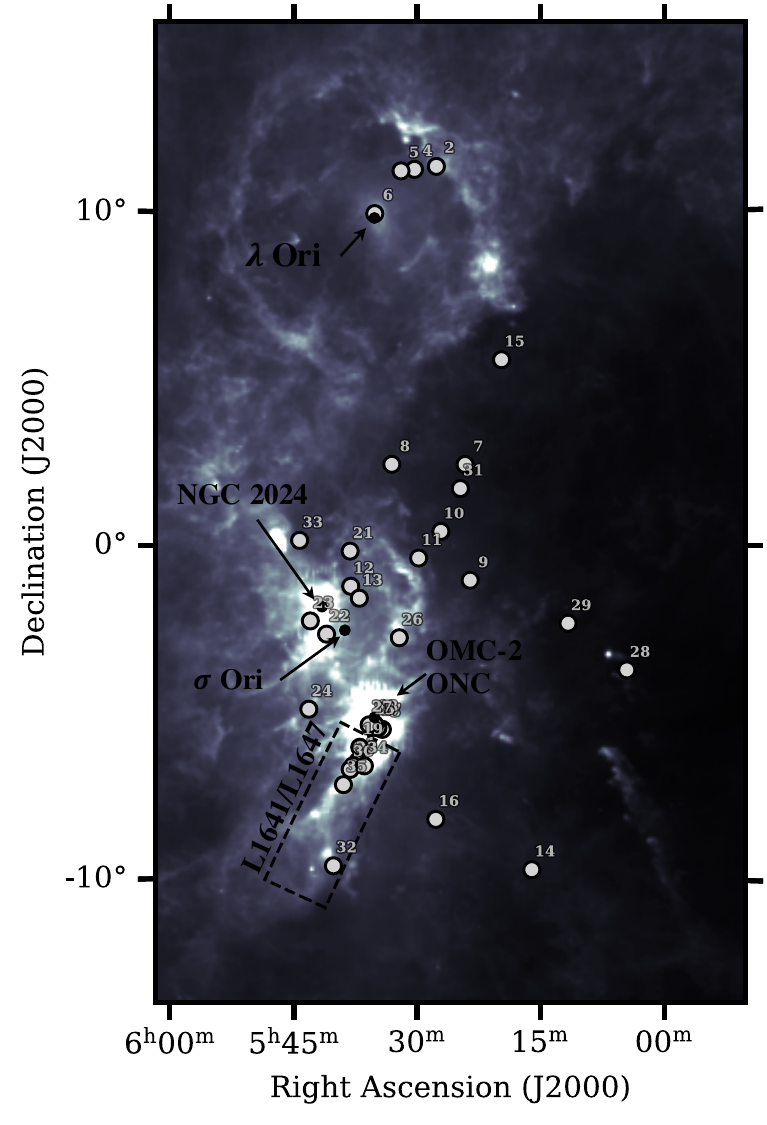}
    \caption{Positions on the sky of the Herbig disk sample used in this work shown as the gray markers. The numbers next to the markers correspond to the numbers in Table~\ref{tab:disk_parameters}. The background is an IRAS 100$\mu$m image \citep{Neugebauer1984}. The positions of six disk surveys have been indicated as well: $\sigma$~Ori \citep{Ansdell2017}, ONC \citep{Eisner2018}, OMC-2 \citep{vanTerwisga2019}, $\lambda$~Ori \citep{Ansdell2020}, NGC~2024 \citep{vanTerwisga2020}, and L1641/L1647 \citep[SODA,][]{vanTerwisga2022}.}
    \label{fig:sky_map_annotated}
\end{figure}

Orion ($\sim$300–475~pc, \citealt{Grossschedl2018}) is the closest star-forming complex with enough pre-main-sequence stars to harbor a sizeable number of Herbig Ae/Be stars. Many population studies have been done in different parts of Orion, tracing different star-formation environments from tranquil regions to more UV-dominated regions. In the $\sigma$ Orionis cluster, an intermediate aged region of 3~Myr old, the disk dust masses were found to depend strongly on the distance to the O9 star at its center \citep{Ansdell2017}. Only 12\% of their disks were more than 10~M$_\oplus$ in mass. Furthermore, CO was only detected in disks more than 1.5~pc separated from the O-star. On the other hand, the disks in the older $\lambda$ Orionis cluster of 5~Myr have not been significantly impacted by the massive stars or the supernova which occurred in the cluster \citep{Ansdell2020}. Interestingly, there is one significant outlier, HD~245185 which is a Herbig disk, which has a more than an order of magnitude higher disk mass than any of the other disks in this region. The younger (0.5~Myr) NGC~2024 cluster was surveyed by \citet{vanTerwisga2020}. This cluster consists of two populations, the eastern population is shielded from FUV irradiation and is similar to other isolated regions, while the western population is older and may be affected by the FUV irradiation. Other younger clusters such as the Orion Nebula Cluster and Orion Molecular Cloud-2 have been surveyed as well \citep{Eisner2018, vanTerwisga2019}, and even class 0 and I surveys have been done \citep{Tobin2020}. Lastly, the largest population study of disks to date, consisting of 873 disks, has been done in the L1641 and L1647 regions of the Orion A cloud (SODA, \citealt{vanTerwisga2022}, for L1641 also see \citealt{Grant2021}).

The plethora of population studies available in Orion gives a solid basis for a comparison between T~Tauri disks and Herbig disks. In this work we present new Northern Extended Millimeter Array (NOEMA) observations of all Herbig disks in Orion. Section~\ref{sec:data_selection_reduction} shows how the targets were selected, the NOEMA data were reduced, and the dust masses were obtained. In Section~\ref{sec:results} the resulting dust mass distribution is presented and compared to distributions of other proto- and pre-main sequence stars in Orion, in addition CO spectra are presented. In Section~\ref{subsec:compare_with_ALMA} the dust mass distribution is compared to the one of ALMA from \citet{Stapper2022}. The impact of external UV irradiation on the dust masses is determined in Section~\ref{subsec:UV}, and a comparison to scattered light is made in Section~\ref{subsec:scattered_light}. Our results are summarized in Section~\ref{sec:conclusion}.

\begin{table*}
\caption{Stellar parameters, $G_0$, measured flux, and derived dust mass of the Orion sample used in this work.}
\tiny\centering
\resizebox{\textwidth}{!}{\begin{tabular}{r|l|rrccccccc}
\hline\hline
\makecell{Nr. \\ \hspace{1mm}} & \makecell{Name \\ \hspace{1mm}} & \makecell{RA \\ (h:m:s)} & \makecell{Dec \\ (deg:m:s)} & \makecell{Dist. \\ (pc)} & \makecell{M$_\star$ \\ (M$_\odot$)} & \makecell{L$_\star$ \\ (L$_\odot$)} & \makecell{Sp.Tp. \\ \hspace{1mm}} & \makecell{Log$_{{10}}(G_{{0}})$ \\ \hspace{1mm}} & \makecell{Flux \\ (mJy)} & \makecell{M$_\text{{dust}}$ \\ (M$_\oplus$)} \\\hline
1. & BF Ori$^{*}$ & 05:37:13.3 & -06:35:01.0 & 378 & 1.9 & 13 & A2 & 1.87 & 0.8 & 1.1±0.1 \\
2. & CO Ori & 05:27:38.3 & +11:25:39.0 & 395 & 2.3 & 23 & F3 & 0.09 & 1.5 & 2.4±0.5 \\
3. & HBC 442 & 05:34:14.2 & -05:36:54.0 & 383 & 2.0 & 10 & F8 & 2.66 & 4.0 & 7.8±1.6 \\
4. & HD 244314 & 05:30:19.0 & +11:20:20.0 & 398 & 2.1 & 19 & A2 & 0.18 & 5.9 & 10.2±2.0 \\
5. & HD 244604 & 05:31:57.3 & +11:17:41.0 & 398 & 2.2 & 34 & A2 & 0.27 & 4.4 & 6.5±1.3 \\
6. & HD 245185$^{*}$ & 05:35:09.6 & +10:01:51.0 & 410 & 2.2 & 30 & B9 & 3.19 & 34.6 & 37.5±3.8 \\
7. & HD 287823 & 05:24:08.0 & +02:27:47.0 & 343 & 1.8 & 12 & A4 & 1.31 & 8.0 & 11.7±2.3 \\
8. & HD 288012 & 05:33:04.8 & +02:28:10.0 & 341 & <1.9 & 14 & A2 & 1.75 & <1.4 & <5.8 \\
9. & HD 290380 & 05:23:31.0 & -01:04:24.0 & 343 & 1.6 & 6 & F5 & 1.09 & 10.6 & 18.6±3.7 \\
10. & HD 290409 & 05:27:05.5 & +00:25:08.0 & 404 & 2.2 & 25 & B9 & 0.90 & 10.1 & 16.8±3.4 \\
11. & HD 290500 & 05:29:48.1 & -00:23:43.0 & 402 & 1.9 & 13 & A2 & 1.45 & 11.9 & 23.7±4.7 \\
12. & HD 290764$^{*}$ & 05:38:05.3 & -01:15:22.0 & 397 & 2.0 & 22 & A5 & 1.71 & 210.1 & 91.2±9.1 \\
13. & HD 290770 & 05:37:02.4 & -01:37:21.0 & 393 & 2.6 & 55 & B8 & 1.70 & 3.5 & 4.4±0.9 \\
14. & HD 34282$^{*}$ & 05:16:00.5 & -09:48:35.0 & 306 & <1.9 & 14 & A0 & 0.05 & 99.0 & 86.7±8.7 \\
15. & HD 34700 & 05:19:41.4 & +05:38:43.0 & 347 & 2.6 & 23 & F8 & 0.02 & 8.1 & 10.2±2.0 \\
16. & HD 35929 & 05:27:42.8 & -08:19:39.0 & 377 & 3.5 & 93 & A9 & 0.73 & <0.7 & <2.0 \\
17. & HD 36917 & 05:34:47.0 & -05:34:15.0 & 445 & 4.4 & 407 & B8 & 2.86 & <0.8 & <2.3 \\
18. & HD 36982 & 05:35:09.8 & -05:27:53.0 & 404 & <6.4 & 1349 & B2 & 3.76 & <0.8 & <1.3 \\
19. & HD 37258$^{*}$ & 05:36:59.3 & -06:09:16.0 & 377 & 2.3 & 26 & A0 & 2.30 & 1.8 & 2.1±0.2 \\
20. & HD 37357$^{*}$ & 05:37:47.1 & -06:42:30.0 & 465 & 2.8 & 87 & B9 & 2.01 & 2.9 & 3.8±0.4 \\
21. & HD 37371 & 05:38:09.9 & -00:11:01.0 & 405 & 3.1 & 100 & B8 & 2.36 & <0.6 & <2.0 \\
22. & HD 37806 & 05:41:02.3 & -02:43:01.0 & 397 & 3.5 & 200 & B8 & 1.90 & 3.4 & 3.1±0.6 \\
23. & HD 38087 & 05:43:00.6 & -02:18:45.0 & 373 & 4.2 & 347 & B5 & 1.79 & <0.6 & <1.2 \\
24. & HD 38120 & 05:43:11.9 & -04:59:50.0 & 381 & 2.8 & 71 & B8 & 0.13 & 31.9 & 35.6±7.1 \\
25. & NV Ori & 05:35:31.4 & -05:33:09.0 & 384 & 2.1 & 59 & F0 & 3.22 & 4.0 & 4.8±1.0 \\
26. & RY Ori & 05:32:09.9 & -02:49:47.0 & 347 & 1.6 & 6 & F4 & 0.64 & 10.6 & 19.1±3.8 \\
27. & T Ori & 05:35:50.5 & -05:28:35.0 & 399 & 2.5 & 59 & A0 & 3.25 & 1.5 & 1.9±0.4 \\
28. & UX Ori & 05:04:30.0 & -03:47:14.0 & 320 & 1.9 & 13 & A3 & 0.29 & 20.2 & 25.2±5.0 \\
29. & V1012 Ori & 05:11:36.5 & -02:22:48.5 & 386 & 1.3 & 6 & A3 & 0.08 & 26.4 & 47.8±4.8 \\
30. & V1787 Ori$^{*}$ & 05:38:09.3 & -06:49:17.0 & 394 & 2.1 & 28 & A3 & 1.47 & 14.8 & 18.2±1.8 \\
31. & V346 Ori & 05:24:42.8 & +01:43:48.0 & 336 & 1.6 & 7 & A7 & 2.32 & 15.3 & 24.8±5.0 \\
32. & V350 Ori & 05:40:11.8 & -09:42:11.0 & 391 & <1.9 & 9 & A1 & 0.27 & 4.4 & 9.1±1.8 \\
33. & V351 Ori & 05:44:18.8 & +00:08:40.0 & 323 & 2.0 & 21 & A7 & 0.18 & 69.4 & 77.4±15.5 \\
34. & V380 Ori$^{*}$ & 05:36:25.4 & -06:42:58.0 & 374 & 2.8 & 95 & B9 & 1.84 & 6.2 & 5.1±0.5 \\
35. & V599 Ori$^{*}$ & 05:38:58.6 & -07:16:46.0 & 401 & 2.1 & 31 & A4 & 0.60 & 55.9 & 69.0±6.9 \\
\hline
\end{tabular}}
\label{tab:disk_parameters}
\tablefoot{For the disks with an asterisk the fluxes have been taken from \citet{Stapper2022, Stapper2024} and \citet{vanTerwisga2022}. In addition to the fluxes there is an absolute calibration error, which is 10\% for ALMA, and 20\% for NOEMA. Except for the UV irradiance $G_0$, continuum flux, and inferred dust mass which we derive in this work, all parameters are taken from \citet{GuzmanDiaz2021}, see this paper for their corresponding uncertainties. For V1012~Ori, the parameters are taken from \citet{Vioque2018}, and the continuum flux is determined from the ALMA pipeline product data. HD~290764 is ALMA Band~7 data.}
\end{table*}

\section{Data selection and reduction}
\label{sec:data_selection_reduction}
In this work we present the results of observations done with the Northern Extended Millimeter Array (NOEMA), with project number S22AU (PI: S. Grant). The disks observed with NOEMA were selected as follows. Using the boundaries of Orion as given in \citet{Zari2017}, we selected all Herbig disks in Orion based on the compilations of \citet{Vioque2018} and \citet{GuzmanDiaz2021}. From this, we obtained a total of 35 Herbig disks located in the Orion star-forming region (see for their positions in Orion Fig~\ref{fig:sky_map_annotated}, the numbers correspond to the numbers in Fig.~\ref{tab:disk_parameters}). Of these 35 disks, ten have existing ALMA interferometric data available (see \citealt{Stapper2022} and \citealt{vanTerwisga2022}, as well as the annotated disks in Table~\ref{tab:disk_parameters}). For one of these ten (V1012~Ori) we obtained publicly available ALMA archival\footnote{\url{https://almascience.eso.org/aq/}} product data (PI: C. Ginski, 2021.1.01705.S). The remaining 25 disks were observed with NOEMA on 2022 December 7 and 30, on 2022 December 17, and on 2023 October 7. Figure~\ref{fig:sky_map_annotated} shows that the targets are spread over the complete constellation. The stellar mass distribution for this sample is not significantly different compared to the all-sky sample of \citet{Stapper2022}, as a Kolmogorov-Smirnov test from \texttt{SciPy} \citep{2020SciPy-NMeth} results in a $p$-value of 0.16.
% In total twelve disks are located in the Orion A molecular cloud and the Orion Nebula Cluster (ONC), while others such as HD~34282, HD~35929, and UX~Ori are isolated.

\begin{figure}[b!]
    \centering
    \includegraphics[width=0.45\textwidth]{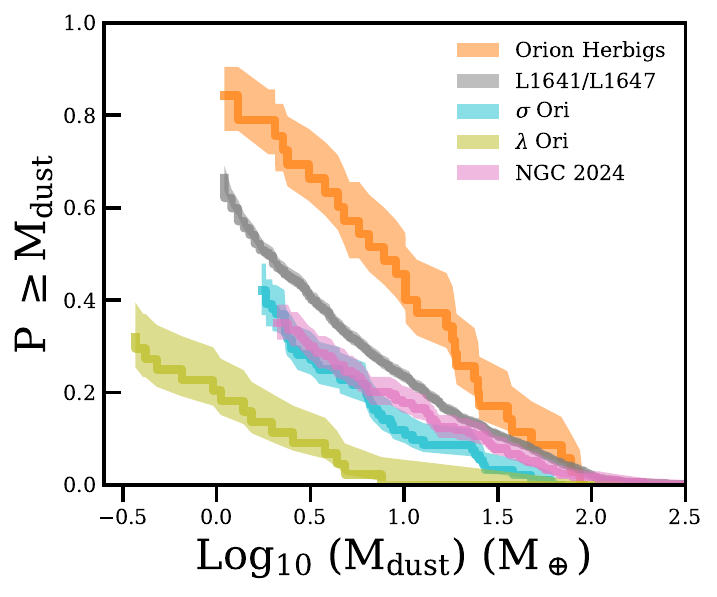}
    \caption{The Orion Herbig disk dust mass distribution compared to other disk surveys done in Orion. The other surveys are: NGC~2024 \citep{vanTerwisga2020},  L1641/L1647 \citep[SODA,][]{vanTerwisga2022}, $\sigma$~Ori \citep{Ansdell2017}, and $\lambda$~Ori \citep{Ansdell2020}.}
    \label{fig:cdf_regions_compare}
\end{figure}

The observations were done in the C configuration. The observations made on 2022 December 17 and November 30 (setup~1) have baselines ranging from 20 to 309 meters with one antenna (\#7) at projected long baselines ranging from 985 to 1397 meters. On 2022 December 7 and 2023 October 7 (setup~2), the baselines range from 20 to 321 meter. Hence, at 210~GHz the largest angular scale the data are sensitive to is $\sim15''$, and a restoring beam of $\sim1''$ is obtained. The wide-band correlator PolyFiX was tuned such that the lower side-band (LSB) and upper side-band (USB) ranged from 203.7 to 211.8~GHz and 219.2 to 227.3~GHz respectively with a channel width of 2~MHz ($\sim2.7$~km~s$^{-1}$). In addition, thirteen high resolution spectral windows with a width of 62.5~kHz ($\sim86$~m~s$^{-1}$) where centered on molecular emission lines such as the $J=2-1$ transition of the CO isotopologues \ce{^13CO}, \ce{C^18O}, and \ce{C^17O}, and other molecules such as \ce{CN} and \ce{H_2CO}.

The NOEMA data were calibrated using the standard pipeline calibration using the \texttt{CLIC} program of the Grenoble Image and Line Data Analysis Software (\texttt{GILDAS}\footnote{\url{http://www.iram.fr/IRAMFR/GILDAS}}). For setup~1 the phase rms threshold was set to 55$\degree$, and the seeing threshold, normally applied to long baseline observations, was not included as only one antenna was at long baselines. For setup~2 the phase rms threshold was set to the default value of 70$\degree$. The calibrators used for setup~1 were 2200+420, LkH$\alpha$~101, 0923+392, and 2010+723 for the passband calibration, and 0458-020 and J0542-0913 were used for the phase and amplitude calibration. All calibrators were used for the flux calibration. The calibrators for setup~2 were LkH$\alpha$~101 and 3C84, which were used for the bandpass calibration, and J0509+056 was used for the phase and amplitude calibration. All calibrators were used for the flux calibration. Due to the reduced phase rms threshold for setup~1, 10\% of the data taken on 2022 November 30 were flagged, while 2\% was flagged for the data taken on 2022 December 17. The weather during observing setup~2 on 2022 November 7 was particularly bad, increasing to a precipitable water vapor of 5~mm by the end of the observations. Due to this, 43\% of the data are flagged. For the observations made on 2023 October 7 nothing was flagged. The on-source integration times ranged from 15 to 30 minutes. After the calibration was done, the data were exported to \texttt{uvfits} files to be further analyzed.

The data were analyzed using the \texttt{Common Astronomy Software Applications} (CASA) application version 5.8.0 \citep{McMullin2007}. To obtain the integrated fluxes in continuum, the LSB and USB were both combined to make a continuum measurement set. This Gaussian is fitted using the \texttt{uvmodelfit} task in CASA after changing the phase-center to be on the target using the \texttt{fixvis} task. As all detections are unresolved, fitting a Gaussian to the visibilities is therefore a good approximation of the observations. From this fit the total flux is obtained. For the non-detections, assuming the emission is coming from a singe beam, three times the rms noise from the empty image was used. For the remaining ten Herbig disks with existing ALMA data, we use the published fluxes as presented in \citet{Stapper2022} and \citet{vanTerwisga2022}, and for V1012~Ori we apply the same method as \citet{Stapper2022} on ALMA archive product data.

To obtain the dust masses, we use the same assumptions as other works \citep[e.g.,][]{Ansdell2016, Cazzoletti2019, Stapper2022}. The flux and dust mass can be related via

\begin{equation}
    M_\text{dust} = \frac{F_\nu d^2}{\kappa_\nu B_\nu(T_\text{dust})},
    \label{eq:Mdust}
\end{equation}

\noindent under the assumption of optical thin emission \citep{Hildebrand1983}. Here, the dust opacity is $\kappa_\nu$, the distance is $d$, and $B_\nu$ is the value of the Planck curve at a dust temperature of $T_\text{dust}$. The dust opacity is estimated as a power-law of the form $\kappa_\nu \propto \nu^\beta$, such that it equals 10~cm$^2$~g$^{-1}$ at a frequency of 1000~GHz \citep{Beckwith1990}. The power-law index $\beta$ is assumed to be equal to 1. To obtain an estimate of the temperature of the dust in Herbig disks we scale the dust temperature $T_\text{dust}$ using the approach of \citet{Andrews2013} via 

\begin{equation}
    T_\text{dust} = 25 \text{ K} \times \left(\frac{L_\star}{L_\odot}\right)^{1/4}.
    \label{eq:Tdust}
\end{equation}

\noindent For the error on the dust masses we use the recommended absolute flux calibration error of 20\% for NOEMA, and 10\% for ALMA. Nevertheless, the dust masses are calculated under the assumption of optically thin emission, which is likely the largest source of uncertainty.

\section{Results}
\label{sec:results}
\subsection{Dust mass distribution}
\label{subsec:dust_mass_distr}

% \subsubsection{Comparison to previous Herbig disk distribution}
Using the \texttt{lifelines} package \citep{DavidsonPilon2021} we obtain the cumulative distribution of the Herbig disks in Orion following \citet{Stapper2022}. The cumulative distribution is presented in Figure~\ref{fig:cdf_regions_compare}, where it is compared to the cumulative dust mass distributions of four regions in Orion (see Fig.~\ref{fig:sky_map_annotated} for their positions in Orion): NGC~2024 \citep[0.5~Myr, ][]{vanTerwisga2020}, L1641/L1647 \citep[SODA, 1-3~Myr, ][]{vanTerwisga2022}, $\sigma$~Ori \citep[3~Myr, ][]{Ansdell2017}, and $\lambda$~Ori \citep[5~Myr, ][]{Ansdell2020}. As there are uncertainties in the determination of the dust masses from the continuum flux using eq.~(\ref{eq:Mdust}), a comparison between only the fluxes is shown in Appendix~\ref{app:flux_distributions}, noting that the flux distribution also depends on the dust temperature and therefore stellar luminosity, an effect that is corrected for in the mass distribution of Fig.~\ref{fig:cdf_regions_compare}.

\begin{figure}
    \centering
    \includegraphics[width=0.5\textwidth]{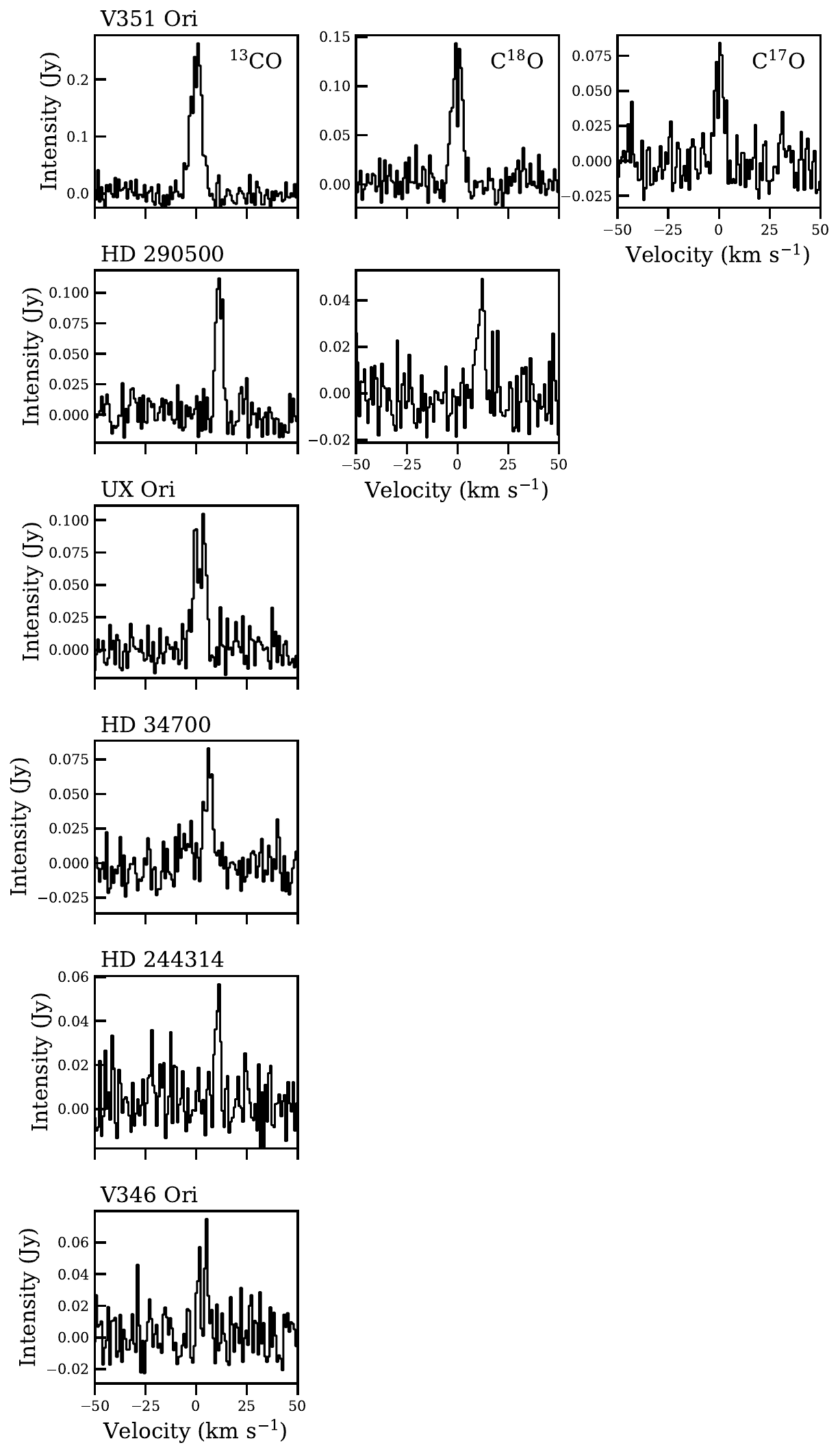}
    \caption{Spectra of the disks in which \ce{^13CO}, \ce{C^18O}, or \ce{C^17O} are detected. The spectra are centered on the frequencies of the emission lines and are binned to 1~km~s$^{-1}$.}
    \label{fig:spectra}
\end{figure}

The dust masses of the Herbig disks range from 91~$M_\oplus$ for HD~290764 down to $<1.2$~$M_\oplus$ for HD~38087. For HD~288012, the disk around the secondary is detected with a flux of 16.5~mJy, corresponding to a disk mass of 65~$M_\oplus$, while for the disk around the Herbig star an upper limit of $<1.4$~$M_\oplus$ is found. The median of the disk masses is 11.7~$M_\oplus$ with a standard deviation of 26~$M_\oplus$, excluding the upper limits.

As noted in \citet{Stapper2022}, we also find that Herbig disks are more massive than T~Tauri disks. As Fig.~\ref{fig:cdf_regions_compare} shows, while the most massive disks are not necessarily part of the Herbig disk population, the number of more massive disks, i.e., disks with a mass more than 10~M$_\oplus$, is at around 50\% for the Herbig disks. For the other regions this is $\sim25$\% or lower.

% \begin{figure*}[t]
%     \centering
%     \includegraphics[width=\textwidth]{cdf_herbig_comparison.pdf}
%     \caption{Comparison of the Herbig sample in Orion of this work with the Herbig disks analyzed in \citet{Stapper2022} and \citet{Stapper2024}. The left panel present the cumulative distributions. The right panel shows the fitted log-normal distributions.}
%     \label{fig:cdf_herbig_compare}
% \end{figure*}

\begin{figure*}[t]
\sidecaption
  \includegraphics[width=12cm]{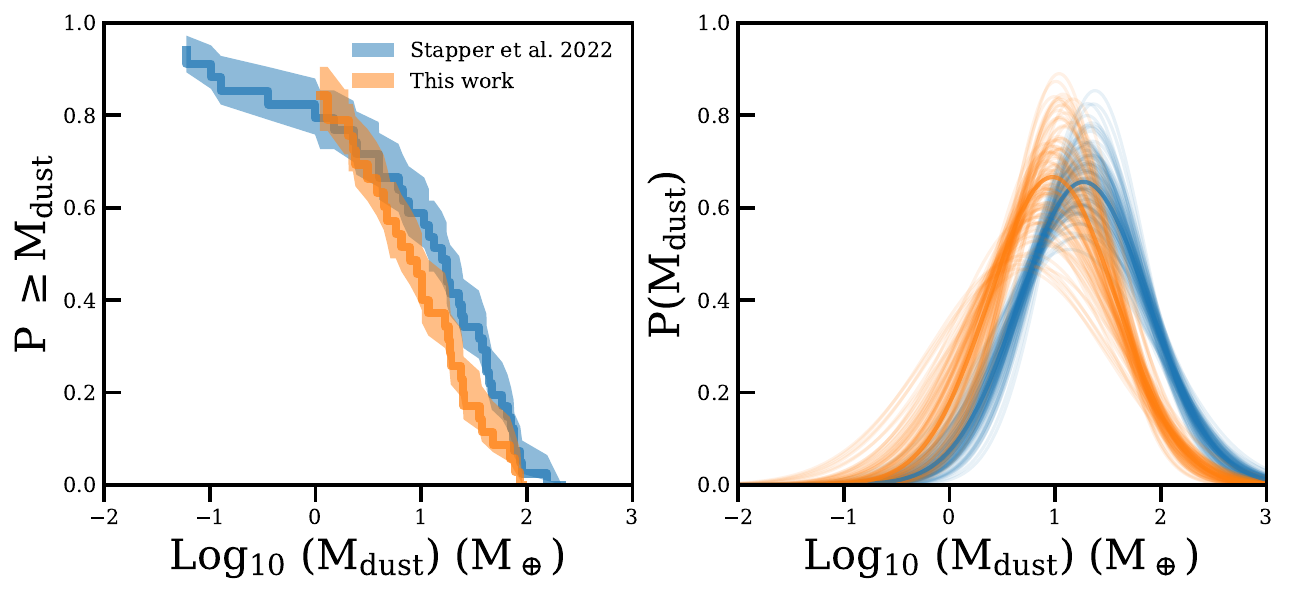}
     \caption{Comparison of the Herbig sample in Orion of this work with the Herbig disks analyzed in \citet{Stapper2022} and \citet{Stapper2024}. The left panel present the cumulative distributions. The right panel shows the fitted log-normal distributions.}
     \label{fig:cdf_herbig_compare}
\end{figure*}

This difference in mass is especially evident when comparing the Herbig disk ages to the ages of the surveyed regions. Based on isochrones from pre-main sequence evolutionary tracks, the median age of the Herbig disks in Orion, after removing upper limits, is 5.1±4.1~Myr \citep{GuzmanDiaz2021}. The ages range from a minimum age of 0.5~Myr to a maximum age of 17~Myr. NGC~2024, the youngest region, is less massive than the Herbig disks; $\sim20$\% of the population is more massive than $10~M_\oplus$, see Fig.~\ref{fig:cdf_regions_compare}. Even splitting the NGC~2024 region in its two populations (the younger east, and older west population), the younger population  has a dust mass distribution up to $\sim30$\% at 10~M$_\oplus$, still lower than what is found for the Herbig disks. 

The difference is even more stark when comparing the dust masses of the oldest region, $\lambda$~Ori, to those of the Herbig disks. All but one disk has a mass below 10~$M_\oplus$. This disk is HD~245185, which is in the 90th percentile of our dust masses\footnote{We adopt a lower disk mass compared to \citet{Ansdell2020} by using a higher disk temperature, 59~K instead of 20~K.}. Regions impacted by external irradiation, such as the disks in $\sigma$~Ori \citep{Ansdell2017, Mauco2023}, show lower disk masses compared to the Herbig disks as well.

% In the Orion A and B clouds the class~0, I and flat spectrum objects have been part of the VANDAM survey \citep{Tobin2020}, allowing for a comparison to the Herbig disks. This comparison is shown in the right panel of Fig.~\ref{fig:cdf_regions_compare}. The dust masses were determined by scaling the dust temperature by the bolometric luminosity of the objects, similar to eq.~(\ref{eq:Tdust}) but with 43~K at 1~$L_\odot$, instead of 25~K. \citet{Tobin2020} found median dust masses of 25.7, 15.6, and 13.8 $M_\oplus$ for the class~0, I, and flat spectrum objects respectively. Though lower disk masses are found when including the disk temperature from radiative transfer \citep{Sheehan2022}. The Class~0 objects are clearly more massive than the Herbig disks, with the median disk mass a factor of $\sim2.5$ higher than for the Herbig disks. The Orion Flat spectrum objects have a remarkably similar dust mass distribution to the Herbig disks.  

\subsection{Gas observations}
The NOEMA observations also covered the \ce{^13CO}, \ce{C^18O}, and \ce{C^17O} $J=2-1$ emission lines. To obtain the spectra of these CO isotopologues, first the phase center was aligned on-target using the \texttt{fixvis} task, after which the \texttt{plotms} task was used to export the measured visibility spectra averaged over time, baseline, and scans. The spectra have been binned by a factor of ten, resulting in a $\sim0.9$~km~s$^{-1}$ resolution.

Figure~\ref{fig:spectra} presents the detected \ce{^13CO}, \ce{C^18O}, and \ce{C^17O} spectra of six disks. For these six disks we detect \ce{^13CO} emission at a peak signal-to-noise of 6 and higher. For only two disks we also detect \ce{C^18O}. In V351~Ori, even \ce{C^17O} is detected. For an additional two disks, T~Ori and HD~36982, foreground cloud emission in the form of large scale emission is detected.

\section{Discussion}
\label{sec:discussion}
\subsection{Comparison to the ALMA Herbig disks}
\label{subsec:compare_with_ALMA}
Figure~\ref{fig:cdf_herbig_compare} compares the obtained cumulative distribution of the Herbig disks in Orion to the distribution of the with all-sky survey of Herbig disks with ALMA of \citet{Stapper2022} combined with the extra sources observed with NOEMA from \citet{Stapper2024}. The detection rate of the Herbig disks in Orion is similar to what was found for the sample analyzed by \citet{Stapper2022}. Out of the 35 disks in the Orion sample, 6 are not detected, resulting in a 83\% detection rate. For the sample of \citet{Stapper2022} 2 out of 36 disks were not detected. However, the ALMA data are more sensitive, and at the NOEMA sensitivity the same detection rate of 83\% would have been obtained for the sample of \citet{Stapper2022}. Indeed, the cumulative distributions shown in Fig.~\ref{fig:cdf_herbig_compare} have the same completeness at a disk mass of $\sim1$~M$_\oplus$.

The main difference between the two dust mass cumulative distributions is at the high end of the distribution. Mostly the higher mass disks are missing in the Orion sample when compared to the sample of \citet{Stapper2022}. To test whether the dust distributions come from a the same population, we use two tests, the \texttt{lifelines} \texttt{logrank\_test} and the Kolmogorov-Smirnov test from \texttt{SciPy} \citep{2020SciPy-NMeth}. These tests test if the Orion distribution is different from the all-sky ALMA distribution, the logrank test also takes upper limits into account. We find $p$-values for the distributions of $p$=0.26 (logrank) and $p$=0.32 (KS), and therefore cannot reject the null hypothesis of both distributions being sampled from the same distribution. However, in the Orion population, massive disks such as HD~97048 and HD~142527 (156~$M_\oplus$ and 215~$M_\oplus$ respectively, \citealt{Stapper2022}) are missing. This suggests that the ALMA all-sky coverage may have been slightly biased toward including the most massive disks, and shows that a complete all-sky survey is warranted.

A lognormal distribution is fitted through the cumulative distributions, following previous works \citep{Williams2019, Stapper2022, Stapper2024}, to obtain probability density distributions. These distributions are shown in the right panel of Fig.~\ref{fig:cdf_herbig_compare}. The mean value of the distributions are Log$_{10}$($M_\text{dust}(M_\oplus)$)=0.94$^{+0.06}_{-0.07}$ and 1.27$^{+0.05}_{-0.05}$ for the Orion sample and the sample of \citet{Stapper2022} respectively. The width of the distributions are the same, 0.59$^{+0.07}_{-0.06}$ and 0.61$^{+0.06}_{-0.06}$ respectively.  There is a clear overlap between the two distributions, further substantiating that the distributions are not significantly different.

Summarizing, we find no significant difference in the distribution of the dust masses of the Herbig disks in Orion compared to the distribution of \citet{Stapper2022}.

\subsection{Impact of UV on Herbig disk masses}
\label{subsec:UV}

\begin{figure}[t!]
    \centering
    \includegraphics[width=0.45\textwidth]{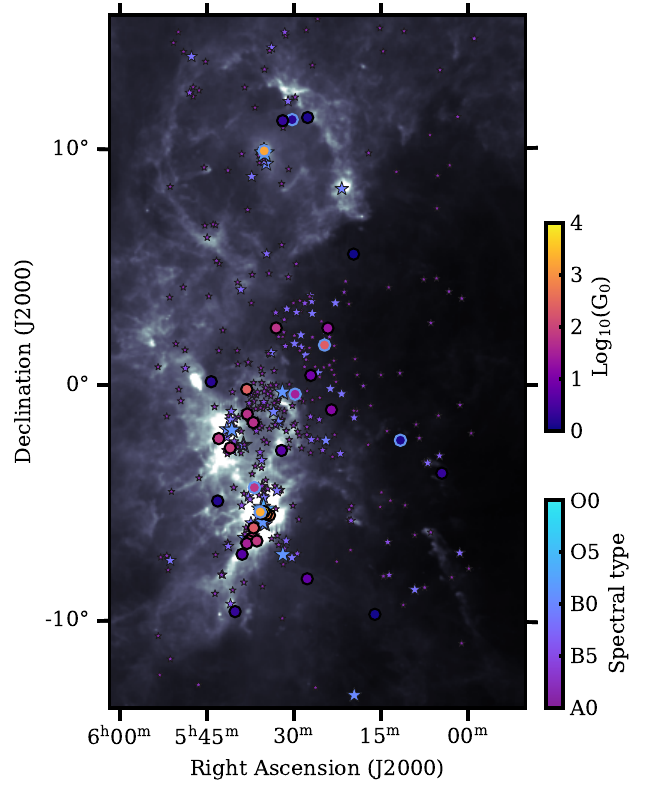}
    \caption{Similar to Fig.~\ref{fig:sky_map_annotated}, but now the FUV irradiation in $G_0$ is shown as the color of the Herbig disk markers. O and B type stars are plotted as the stars, where the color (and size) indicates their spectral type. The Herbig disk with a blue edge have also been detected with NOEMA in at least \ce{^13CO}. For HD~34282 \ce{^13CO} and \ce{C^18O} have been detected with ALMA, and for HD~245185 and V1012~Ori \ce{^12CO} is very bright and detected with ALMA \citep{Stapper2024}.}
    \label{fig:sky_map_G0}
\end{figure}

Orion contains a large number of massive young stars which contribute to the far ultraviolet (FUV) external irradiation field, which influences the amount of mass present in the disk by triggering photoevaporative winds. Figure~\ref{fig:sky_map_G0} shows the external ionizing sources as the stars and the corresponding FUV impact on the Herbig disks as the purple to yellow shaded circles in terms of $G_0$ ($1.6\times10^{-3}$ erg~cm$^{-2}$~s$^{-1}$, \citealt{Habing1968}). The Herbig disks with a blue edge have also been detected with NOEMA in at least \ce{^13CO}. For HD~34282 \ce{^13CO} and \ce{C^18O} have been detected with ALMA \citep{Stapper2024}, and for HD~245185 and V1012~Ori \ce{^12CO} is very bright and detected with ALMA \citep{Stapper2024}.

\begin{figure}[t!]
    \centering
    \includegraphics[width=0.45\textwidth]{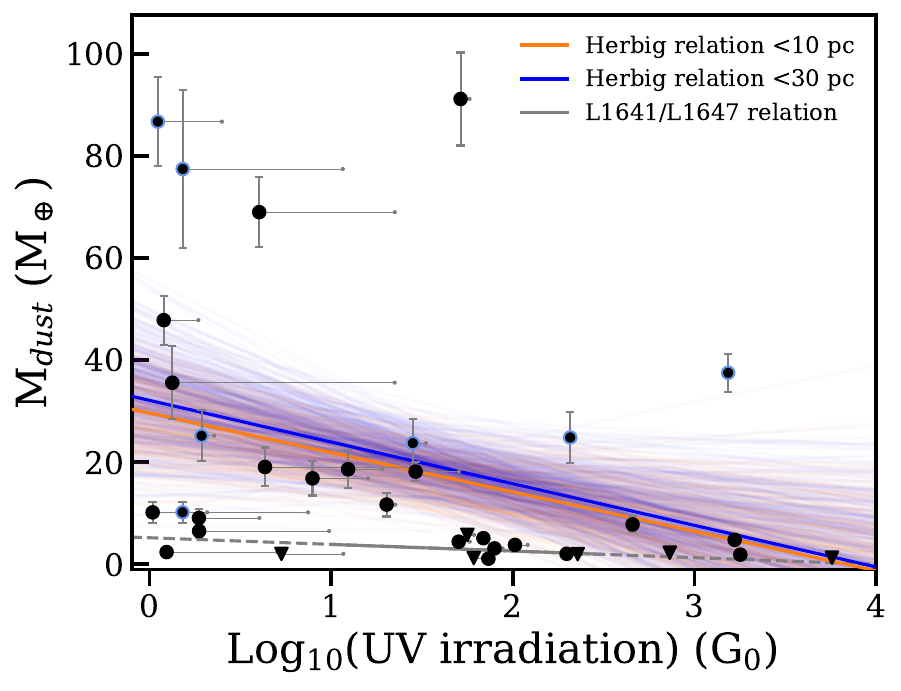}
    \caption{The external UV irradiation plotted against the dust mass of the Herbig disks. Ionizing stars within 10~pc are have been taken into account, a relationship is fitted through these data in orange. The gray horizontal lines indicate where the points move if a cutoff of 30~pc is used, and a corresponding fit is done in blue. The blue marker outlines indicate disks in which CO is detected (see Fig.~\ref{fig:spectra}, or \citealt{Stapper2024}). The relationship found for L1641/L1647 is shown as the gray line \citep{vanTerwisga2023}. The region over which this line was fitted is solid, while the dashed line is extrapolated.}
    \label{fig:G0_Mdust}
\end{figure}

The ionizing stars were found, following \citet{vanTerwisga2023}, by querying \texttt{Simbad}\footnote{\url{http://simbad.u-strasbg.fr/simbad/}} for stars with spectral types earlier than A0 within the region on the sky as shown in Figs.~\ref{fig:sky_map_annotated} and \ref{fig:sky_map_G0}. The distances to the stars were limited to 300-475~pc, and as the uncertainties on these distances can be fairly large only projected distances are used. The FUV luminosities of these stars were then computed by using BHAC-15 isochrones \citep{Baraffe2015} and integrating model spectra \citep{Castelli2003} between 911.6 and 2066~Å. As was done by \citet{vanTerwisga2023}, no stars beyond a projected distance of 10~pc from a given Herbig disk were taken into account, but interstellar extinction was not included otherwise. Including irradiation of stars out to 30~pc primarily changes the $G_0$ of the least irradiated disks to at most a few 10s~$G_0$. The minimum UV irradiation was set at 1~$G_0$. The resulting values of $G_0$ are shown in Fig.~\ref{fig:sky_map_G0}, and compared to the dust masses in Fig.~\ref{fig:G0_Mdust}; disks with CO detections are indicated as in Fig.~\ref{fig:sky_map_G0}.

% Figure~\ref{fig:G0_Mdust} presents the dust mass as a function of UV irradiation. By binning the dust masses in logarithmically spaced bins in UV irradiation, we can assess whether there is a trend between the two variables. The solid black line is using the mean value of each bin, the dashed black line is using the median per bin. The gray line in Fig.~\ref{fig:G0_Mdust} is the relationship between the dust mass and UV irradiation found for L1641/L1647 \citep{vanTerwisga2023}.

% Depending on the distance cutoff we adopt to include contributions to Fuv (10 pc or 30 pc), we find a relation M_d = a*log(Fuv) +b with respective slopes of -7.6+/-3.5 and -8.5+/-4.5, and offsets of +29.4+/-6.7 and +32.8+/-8.7. Although the slopes appear steeper than found by -1.3+0.14-0.13 by vT03 for L1641/L1647, they agree with the latter value within 2sigma. The findings support a trend of decreasing dust mass with increase UV irradiation.

Based on Fig.~\ref{fig:G0_Mdust}, there is a trend with a decrease in dust mass for an increase in UV irradiation for Herbig disks. We use the \texttt{linmix}\footnote{\url{https://linmix.readthedocs.io/}} package \citep{Kelly2007} to fit a linear relationship through the data. This fit results in the orange and blue relationships shown in Fig.~\ref{fig:G0_Mdust}. Depending on the distance cutoff we adopt to include contributions to $F_\text{uv}$ (10~pc or 30~pc), we find a relation $M_\text{dust}=a\times \log (F_\text{uv}) + b$ with respective slopes of -7.6±3.6 and -8.5±4.5, and offsets of 29.4±6.7 and 32.8±8.7. Although the slopes appear steeper than $-1.3^{+0.14}_{-0.13}$ for L1641/L1647 \citep{vanTerwisga2023}, they agree with the latter value within $2\sigma$. The findings support a trend of decreasing dust mass with increase UV irradiation.

% We find the relationship $M_\text{dust}=-7.6\pm3.6 \times$~log$_{10}(F_\text{uv}/G_0)+29.4\pm6.7$, corresponding to a decrease in dust mass with an increase in $G_0$. This relationship is steeper compared to that of L1641/L1647 which has a slope of $-1.3^{+0.14}_{-0.13}$ \citep{vanTerwisga2023}, but is also consistent within $2\sigma$ given the errors on the relationships. Additionally, including ionizing stars out to 30~pc the relationship becomes $M_\text{dust}=-8.5\pm4.5 \times$~log$_{10}(F_\text{uv}/G_0)+32.8\pm8.7$, slightly steeper compared to the cutoff of 10~pc, see Fig.~\ref{fig:G0_Mdust}.

% Especially the mean binned dust masses clearly decrease with an increase in UV irradiance. Using the median disk mass to remove the influence of outliers, the trend is still visible. Fitting a linear relationship through the binned data using the \texttt{curve\_fit} routine from \texttt{SciPy} results in the red and orange relationships shown in Fig.~\ref{fig:G0_Mdust}. For the mean dust masses we find $M_\text{dust}=-7.2\pm1.1 \times$~log$_{10}(F_\text{uv}/G_0)+28.0\pm2.5$, and for the median dust masses we find $M_\text{dust}=-5.2\pm1.0 \times$~log$_{10}(F_\text{uv}/G_0)+18.9\pm2.3$. We find a steeper relationship compared to that of L1641/L1647 which has a slope of $-1.3^{+0.14}_{-0.13}$ \citep{vanTerwisga2023}.

There are disks with dust masses higher than 10~M$_\oplus$ for all bins, except at the highest UV irradiation. Furthermore, CO is detected over all UV irradiation values traced, which is a tracer of the UV irradiance as well. For some of the disks with the strongest UV irradiance, we still clearly detect CO emission. The UV irradiation increases the temperature of the disk both in the dust and the gas, which is used to explain the bright CO emission seen in irradiated T~Tauri disks due to less freeze-out \citep{Boyden2020, Ballering2023}. But for Herbig disks the star dominates: using eq.~(2) from \citealt{Ballering2023}, the Herbig disk needs to be within (a projected distance of) $\sim0.3$~pc of an O-star (which none of our disks are) for the temperature set by UV irradiation to dominate over the temperature set by the stellar luminosity (see eq.~(\ref{eq:Tdust})). Especially HD~245185 is of note, for which strong \ce{^12CO} emission has been detected with ALMA \citep{Ansdell2020}, even though it has one of the highest irradiances of the sample used in our work. This might be indicative of a chance alignment, and that the Herbig disk is actually further away from the O-star than it seems. Placing HD~245185 at the same distance as the shell seen around $\lambda$~Ori (Figs.~\ref{fig:sky_map_annotated}, \ref{fig:sky_map_G0}), results in a distance of $\sim20$~pc, instead of less than 1~pc. This would decrease the irradiance to negligible values of around 2-4~$G_0$, aligning the disk mass with the seen trend. The other high disk masses might therefore also be chance alignments. Still, disks likely need to be close to the O-star for their chemistry to be affected by UV irradiation \citep{RamirezTannus2023, DiazBerrios2024}.

% \textbf{Lastly, we note that there are T~Tauri disks in $\sigma$~Ori which have masses of more than 10~M$_\oplus$ with irradiation levels of $10^3$~$G_0$ \citep{Mauco2023}. Similar mass Herbig disks can be found at these irradiation levels as well. Hence, this comparison could very much depend on the region.}
% Still, the trend does seem to be steeper when compared to T~Tauri disks \citep{vanTerwisga2023}. This is expected from disk modeling, as the Herbig stars are more massive than T~Tauri stars, the materials in the disk are stronger bound and thus a higher irradiance is necessary to start photo-evaporating the disk. Hence, external UV irradiation is not expected to be as impactful on Herbig disks as it would be for disks around T~Tauri stars \citep[e.g.,][]{Haworth2018, Haworth2023}.

% \subsection{Completeness of the Herbig sample}

\subsection{Comparison to scattered-light data}
\label{subsec:scattered_light}
Recently \citet{Valegard2024} published a survey of disks in Orion observed in scattered light imaging with SPHERE/VLT. This survey was done as part of the DESTINYS program (recent papers include \citealt{Garufi2024, Ginski2024, Valegard2024}). The survey consists of 23 stars in Orion with spectral types ranging from A0 to K6. Out of the 23 disks 10 have a detection of a disk, and four of these have a clearly resolved disk.

Out to Orion only few disks are possible to resolve with SPHERE. In the sample of \citet{Valegard2024}, four disks are resolved: HD~294260, V1012~Ori, V351~Ori, and V599~Ori. Out of these four disks, three are Herbig disks and some of the more massive disks in our sample. HD~294260 is a intermediate mass T~Tauri (IMTT) star, which are precursors of Herbig stars \citep{Valegard2021}. The disk around IMTT stars show similar characteristics as the disks around Herbig stars, with the same dust mass distribution \citep{Stapper2024_IMTT}. The dust mass of HD~294260 is found to be 74.4±7.9~M$_\oplus$ \citep{Stapper2024_IMTT}. Other disks resolved with scattered-light imaging in Orion include HD~34282 \citep{deBoer2021} and HD~290764 \citep{Ohta2016}. All of these disks are around intermediate mass stars. 

Some of these disks also show clear signs of asymmetries in the polarized scattered light. V351~Ori shows multiple asymmetric structures \citep{Wagner2020, Valegard2024}, in particular arc-like structures in the outer ring of the disk. V599~Ori has indications of a spiral arm, the inner disk is brighter in the southeast and the outer disk in the northwest \citep{Valegard2024}. These types of structures are generally explained by giant planets residing in these disks \citep[see for an overview][]{Bae2023}. As around intermediate mass stars giant exoplanet formation is highest \citep[e.g.,][]{Johnson2007, Johnson2010, Nielsen2019}, the fact the massive and largest disks with substructures found around intermediate mass disks could be related to this. As \citet{Stapper2022} and others \citep[e.g.,][]{Maaskant2013} proposed, the evolution of Herbig disks is likely significantly influenced by these giant exoplanets, which can keep the disk large and bright. A similar hypothesis was put forward by \citet{Ansdell2020} for the HD~245185 disk, which is by an order of magnitude the most massive disk in $\lambda$~Orionis. The host star of the HD~245185 disk was found to also be depleted in refractory elements, suggesting giant planet formation occurring in this disk \citep{Kama2015, GuzmanDiaz2023}. Similarly, V599~Ori was also found to be depleted in refractory elements \citep{GuzmanDiaz2023}. While these large disks are not lacking around T~Tauri disks (e.g., V1094~Sco \citealt{vanTerwisga2018}, IM~Lup \citep{Andrews2018b}, Sz~98 \citealt{Ansdell2016}), they do seem to be more common around Herbig disks.

\section{Conclusion}
\label{sec:conclusion}
In this paper we present the first complete survey of Herbig disks in a single region. We present new NOEMA observations of 25 Herbig disks which, together with 10 archival ALMA observations, cover all Herbig disks in Orion. Based on these observations we conclude the following:

\begin{enumerate}
    \item The Herbig disks in Orion are found to have a median dust disk mass of 11.7~$M_\oplus$, ranging from 91~$M_\oplus$ down to an upper limit of $<1.2$~$M_\oplus$.
    \item Comparing the Herbig disks in Orion to previous surveys done in Orion, we find a higher mean dust disk mass compared to the T~Tauri disks. While for Herbig disks 50\% of the disks have masses higher than 10~M$_\oplus$, for T~Tauri disks this is 25\% or lower. This difference is especially apparent when considering that the Herbig disks have a median age of 5.1~Myr, while the star-forming regions in Orion are as young as 0.5~Myr.
    \item There are no significant differences between the dust mass distribution of the Orion Herbig disk population and those observed with ALMA across the sky \citep{Stapper2022}. The only difference appears due to the lack of a few individual objects with particularly large disk masses (beyond $\sim150$~$M_\oplus$), that are absent from the Orion sample.
    \item Herbig disks show a steeper trend between disk dust mass and UV irradiation compared to T~Tauri disks. A slope of -7.6 is found, compared to -1.3 for T~Tauri disks.
    \item The largest disks in the recent SPHERE survey of disks in Orion of \citet{Valegard2024} are Herbig disks and some of the most massive disks in our sample, likely relating to giant exoplanet formation occurring in these disks.
\end{enumerate}

This work has shown the importance of complete studies of Herbig disks. As we are going towards more complete and better defined samples of Herbig stars, we should push for complete millimeter observations of the Herbig disk population.

\begin{acknowledgements}
The research of LMS is supported by the Netherlands Research School for Astronomy (NOVA). This work is based on observations carried out under project number S22AU with the IRAM NOEMA Interferometer. IRAM is supported by INSU/CNRS (France), MPG (Germany) and IGN (Spain). We would like to thank Jan Orkisz as our local contact at IRAM. ALMA is a partnership of ESO (representing its member states), NSF (USA) and NINS (Japan), together with NRC (Canada), MOST and ASIAA (Taiwan), and KASI (Republic of Korea), in cooperation with the Republic of Chile. The Joint ALMA Observatory is operated by ESO, AUI/NRAO and NAOJ. This work makes use of the following software: The Common Astronomy Software Applications (CASA) package \citep{McMullin2007}, Python version 3.9, astropy \citep{astropy2013, astropy2018}, lifelines \citep{DavidsonPilon2021}, matplotlib \citep{Hunter2007}, numpy \citep{Harris2020}, and scipy \citep{2020SciPy-NMeth}. Lastly, we thank the referee for their careful consideration of our work and for their thoughtful comments which improved the manuscript.

\end{acknowledgements}

\bibliographystyle{aa}
\bibliography{references.bib}

\appendix
\section{Flux distributions}
\label{app:flux_distributions}
\begin{figure}[h!]
    \centering
    \includegraphics[width=0.5\textwidth]{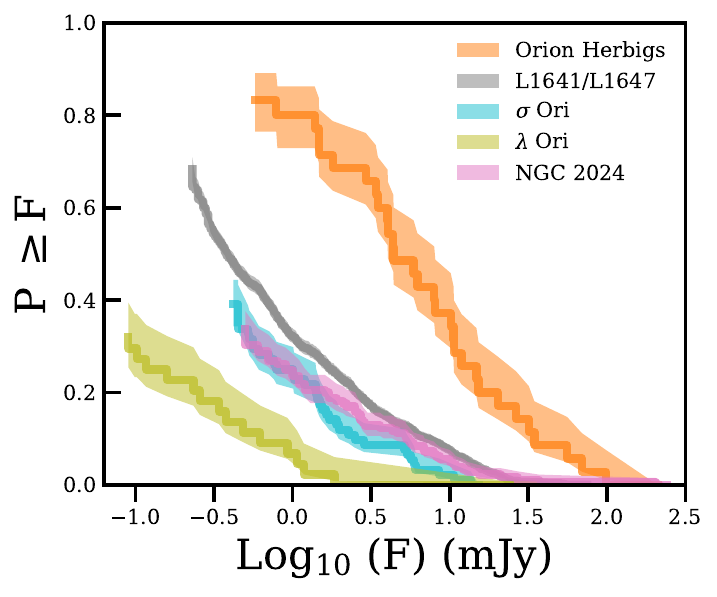}
    \caption{Same figure as Figure~\ref{fig:cdf_regions_compare}, but now for the fluxes, not converted to dust mass.}
    \label{fig:cdf_fluxes}
\end{figure}

Figure~\ref{fig:cdf_fluxes} presents cumulative distributions of the same regions as shown in Figure~\ref{fig:cdf_regions_compare}, but now for the observed fluxes, instead of the dust masses as converted via eq.~(\ref{eq:Mdust}). Due to the relatively warm disks around Herbig stars, the Herbig disks now stand out more prominently compared to the other regions.

% stand out more prominently

\end{document}